# Connectivity Compression for Irregular Quadrilateral Meshes

Davis King, Jarek Rossignac, and Andrzej Szmczak
kingd@cc.gatech.edu, jarek@gvu.gatech.edu, andrzej@cc.gatech.edu,
Graphics, Visualization, and Usability Center, College of Computing
Georgia Institute of Technology

## *Abstract*

Applications that require Internet access to remote 3D datasets are often limited by the storage costs of 3D models. Several compression methods are available to address these limits for objects represented by triangle meshes. Many CAD and VRML models, however, are represented as quadrilateral meshes or mixed triangle/quadrilateral meshes, and these models may also require compression. We present an algorithm for compressing such quadrilateral meshes, and we demonstrate that in general they may be encoded with fewer bits than triangle meshes with the same number of vertices. By preserving and exploiting the original quad structure, our approach achieves encodings 30-80% smaller than an approach based on randomly splitting quads into triangles. We present both a code with a proven worst-case cost of 2.67 bits per vertex for meshes without valence-two vertices and entropy-coding results for typical meshes ranging from 0.3 to 0.9 bits per vertex, depending on the regularity of the mesh. Our method may be implemented by a rule for a particular splitting of quads into triangles and by using the compression and decompression algorithms introduced in [Rossignac99] and [Rossignac&Szymczak99]. We also present extensions to the algorithm to compress meshes with holes and handles and meshes containing triangles and other polygons as well as quads.

# Motivation

In many applications, it is necessary to transmit 3D models over the Internet -- to share CAD/CAM models with e-commerce customers, to update content for entertainment applications, or to support collaborative design, analysis, and display of engineering, medical, and scientific datasets. Bandwidth constraints and storage costs, however, limit the complexity of the 3D models that can be used over the Internet and other networked environments.

A common representation for 3D objects is an irregular mesh of polygons whose union defines a surface bounding a solid object. A simple representation for a polygon mesh stores *the geometry* (a table of the coordinates of its V vertices) and *the connectivity* (a table of its F faces, each represented by a list of vertex indices of at least $\log_2(V)$ bits each and by the number of vertices in the list[*]). Previously reported compression methods for the vertex coordinates (geometry) have used vertex quantization, geometric predictors, and variable-length encodings of corrective vectors to compress the vertex coordinates down to 4 or 5 bits each for a class of models [Deering95, Chow97, Hoppe98, Taubin&Rossignac98]. Since the uncompressed representation of the connectivity, as described above, requires $O(V \log V)$ space, it dominates storage for large V, and it is important to develop compact representations for it as well.

Several compressed representations have been developed for the connectivity of irregular triangle meshes [Deering95], [Taubin&Rossignac98], [Rossignac99], [Gumhold&Strasser98], [Touma&Gotsman98], [Taubin&Gueziec98], [Bajaj99], [Isenburg&Snoeyink99], [Li99]. The best of these approaches achieve results less than 2 bits/vertex with entropy coding for some typical 3D models, and guaranteed worst case bounds as low as 3.67 bits/vertex [King&Rossignac99] for any triangle mesh.

---

[*] For a triangle or quadrilateral mesh, it is unnecessary to store this number since each face is represented by a list of three or four vertex indices.

While triangle meshes (T-meshes) are common, many CAD, VRML, and other 3D model libraries (such as Viewpoint's 3D model repository [Viewpoint]) contain quadrilateral meshes (Q-meshes), mixed triangle/quadrilateral meshes (QT-meshes), or meshes including other kinds of polygons.

Some finite-element simulations use irregular Q-meshes and QT-meshes, especially for modeling structures and simulating torsion and crashes [Müller-Hannemann97]. In computer graphics, physically-based modeling systems have used Q-meshes for modeling cloth deformations [Terzopoulos87] and, recently, for simulating weathered stone [Dorsey99]. Some scanners sample points at the vertices of a regular, square lattice, and then combine such vertices into a single irregular mesh. Many curved surface primitives such as B-splines, NURBS, and Catmull-Clark subdivision surfaces use irregular Q-meshes of control points to specify the underlying surface. Mixed QT-meshes often result from joining such curved primitives with triangle-based primitives and from using triangles to repair gaps between B-spline patches. Furthermore, some users may prefer quadrilaterals to triangles for aesthetic reasons or ease of modeling [DeRose98].

Several authors provide further background on the issues that arise in the use of Q-meshes and QT-meshes for CAD modeling [Nowottny99], [Weihe99], [Müller-Hannemann97]. Nowottny, Bern and Epstein, and Müller-Hannemann also present algorithms for generating irregular Q-meshes [Nowottny99], [Bern&Epstein97], [Müller-Hannemann97].

One simple way to compress a quadrilateral mesh or a mesh containing other polygons is to split each polygon into triangles and to apply one of the triangle mesh compression algorithms. To maintain compatibility with the original mesh, decompression may need to delete the added edges and restore the original quads. As a result, compressing these meshes by triangulating all polygons may require encoding additional information describing how to recover the original polygon mesh structure from the stream of compressed triangles.

Instead, our results demonstrate that it is more efficient to exploit the original quad structure than to triangulate a mesh arbitrarily before compression, whether or not it is necessary to preserve the original quads.

## Summary of Our Results

This paper makes both a practical contribution as the first compression algorithm designed for quad and mixed tri/quad meshes, and a theoretical contribution by showing that both the worst-case and entropy-coding costs of encoding quadrilateral meshes are less than the corresponding costs of encoding triangle meshes with the same number of vertices.

Our approach is a generalization of the linear-time compression (Edgebreaker) and decompression (Wrap&Zip) algorithms introduced for triangle meshes in [Rossignac99] and [Rossignac&Szymczak99]. The essence of our approach is to split each quad into triangles according to a rule ensuring that the two triangles created from each quad are adjacent in Edgebreaker's traversal sequence (a triangle spanning tree). This splitting rule is easy to implement, and it allows the original quads to be recovered by simply joining adjacent pairs of triangles during decompression. It leads to efficient encodings because it makes some combinations of the Edgebreaker labels impossible.

We provide a variable-length code that guarantees to represent any simple Q-mesh with no internal valence-two vertices in 2.67 bits per vertex or less.[*] These codes are practically useful for compression, since they avoid the overhead of constructing different entropy codes for each model. Our worst-case cost of 2.67 bits per vertex is also theoretically significant, since it is known that for every possible triangle mesh compression scheme there are triangle meshes whose encoding requires 3.24 bits per vertex or more [Tutte62] (see also [Keeler&Westbrook95]).

---

[*] A polygon mesh consists of a topological graph of vertices, edges, faces, and vertices and a set of vertex coordinates describing the embedding of the graph in space. A mesh is manifold if and only if each edge is incident to exactly two faces and the faces incident to each vertex form a single cone. A mesh may also be described as manifold with boundary if it meets the conditions for being manifold everywhere except for one

We also present empirical results demonstrating that our approach can achieve even better compression ratios in practice, and that it significantly outperforms an approach based on randomly splitting quads into triangles. On a variety of Q-meshes used in practice, our method costs between 1.3 and 2 bits per vertex with our fixed variable-length code and between 0.3 and 0.9 bits per vertex via entropy coding. Entropy-coded files created with our approach are consistently 30-80% shorter than entropy-coded files created by randomly splitting quads into triangles.

For QT-meshes, we show that our approach can achieve 40-60% improvements via entropy coding for large QT-meshes. This result implies that our methods may be used to compress triangle meshes created from a mix of joined B-spline/NURBS patches and irregular triangles more efficiently than methods that do not consider the original mesh polygons. For the quadrilateral regions of such meshes, our splitting rule is also highly effective at creating long triangle strips for transmission to rendering hardware. We also present a code that guarantees to encode any polygon mesh in 5 bits per vertex or less.

As a further theoretical contribution, we compare the information-theoretic limits on encoding quad meshes to Tutte's limits on triangle meshes [Tutte62]. Our worst-case result of 3.07 bits/vertex for encoding simple Q-meshes (with internal valence-two vertices allowed) implies that the number of possible planar Q-meshes is less than $O(2^{3.07V})$, and we present an additional proof showing that the number of planar Q-meshes is at least $\Omega(2^{2.24V})$. These results confirm our claim that Q-meshes may be compressed more efficiently as Q-meshes than by converting them into T-meshes via random splits, but they also suggest that further gains in Q-mesh encoding are possible.

## Background and Prior Art on Connectivity Compression

Several algorithms have been developed to address the problem of compactly encoding the connectivity of polygonal meshes, both as the theoretical problem of short encodings of embedded graphs and as a practical problem of compressing the incidence table of triangle meshes for 2D and 3D models.

Publications on these algorithms have reported their results in a variety of ways, some reporting the cost in bits per vertex [Turan84] [King&Rossignac99], some in bits per face [Rossignac99, Gumhold&Strasser98], and others in bits per edge [Keeler&Westbrook95]. To compare our results to prior work, we use Euler's formula to derive relationships among the number of vertices, edges, and faces in a Q-mesh and in a QT-mesh. We use these relationships to express both our own results and published data in terms of bits per vertex; the same formulae may also be used to convert our results into bits per edge or bits per face.

Consider a polygon mesh with V vertices, E edges, and F faces. Let Q be the number of quadrilateral faces and let T be the number of triangle faces. For a QT-mesh, F=Q+T, and for a Q-mesh, F=Q. In a mesh without boundary, each edge is incident to 2 faces and is thus used twice. The total number of edge-uses is therefore 2E. Since each quad is incident to 4 edges and each triangle is incident to 3 edges, the total number of edge-uses is also equal to 4Q+3T. Therefore:

E=2Q+3T/2 (1)

---

or more loops of edges incident to only one face each. Such edges are boundary edges, and loops of boundary edges are called holes. A *simple mesh* is a manifold mesh without boundary which is topologically equivalent to a sphere (i.e. deleting any loop of edges will disconnect the mesh into two separate connected components).

For a simple polygon mesh, Euler's formula is V-E+F=2. For a Q-mesh with T=0, E=2Q, and Euler's formula implies that Q=V–2 and E=2V–4. For a QT-mesh, E ranges from 2V–4 to 3V–6, and we have:

$$V=2+Q+T/2 \qquad (2)$$

$$E=2V-4+T/2 \qquad (3)$$

The theoretical problem of succinctly encoding planar graphs has been studied extensively in the graph-theoretic literature [Tutte62, Tutte73, Itai&Rodeh82, Turan84, Naor90, Kao&Teng94, Keeler&Westbrook95, Munro&Raman97, Chuang&others98, He&Kao99]. Most of these theoretical results may be applied to any polygon mesh, and the costs of encoding Q- and QT-meshes with each of these approaches may be deduced from the published worst-case bounds. Using (2) and (3) above, Keeler&Westbrook's [Keeler&Westbrook95] worst-case bound of 3 bits/edge for any polygon mesh is equivalent to 6 bits/vertex for a quad mesh and a value between 6 to 9 bits/vertex for any tri/quad mesh. Chuang [Chuang98] achieves guaranteed encodings of less than 2.5+2log3 bits/vertex or 5.67 bits/vertex for Q- and QT-meshes, and He&Kao [He&Kao99] achieves encodings of less than log3(V+E) bits, or 4.75 bits/vertex for Q-meshes.

While the prior theoretical work discussed above has focused on reducing the worst-case encoding costs of planar graphs, practical algorithms for triangle mesh compression have focused on compressing the incidence tables for triangle meshes used for typical 2D and 3D models. Each of these methods may be used to compress a Q-mesh or QT-mesh by simply splitting the quads into triangles before compression, and, if necessary, marking which pairs of triangles should be joined into quads during decompression.

Deering and Chow [Deering95, Chow97] use generalized triangle strips and a vertex cache to encode the connectivity of a mesh and to minimize the number of times a vertex must be transmitted to rendering hardware. Since this approach is designed for transmitting streaming geometry to rendering hardware, it places a greater priority on the lengths of the resulting triangle strips than on minimizing the overall encoding cost.

Several approaches [Taubin&Rossignac98, Gumhold&Strasser98, Touma&Gotsman98, Rossignac99, Rossignac&Szymczak99, King&Rossignac99, Bajaj99, Isenburg&Snoeyink99] are based on encoding either a triangle-spanning or a vertex-spanning tree. Taubin et al. were the first to demonstrate that these approaches may be generalized to polygons by systematically splitting a polygon mesh into triangles and encoding the resulting triangle mesh with Taubin and Rossignac's Topological Surgery method [Taubin et al. 98]. To preserve the original polygons, their approach marks each interior edge in the spanning tree with a bit indicating whether the edge should be deleted, at a cost of 2 bits per vertex above the cost of encoding the corresponding T-mesh via Topological Surgery.

Bajaj [Bajaj et al. 99] has also developed an extension of their mesh compression algorithm to polygon meshes. However, they have not reported any per-vertex or per-polygon connectivity costs for this extension, either for the worst-case or for typical meshes.

In each of these methods, the spanning tree defines a particular traversal sequence or canonical ordering of the triangles. The efficiency of compressing a quad mesh by splitting it into triangles and using one of these algorithms depends on how the chosen sequence of splits affects the traversal sequence of the mesh. Our approach may be implemented as a particular rule for splitting quads into triangles systematically. While our splitting rule is designed for Edgebreaker, Gumhold & Strasser and Touma & Gotsman's algorithms use the same traversal sequence as Edgebreaker. Their algorithms, therefore, may be adapted to use the same splitting rule to achieve efficient encodings of Q-meshes.

## Our Approach

### *Edgebreaker Compression*

Our approach to encoding Q-meshes and QT-meshes is a generalization of the Edgebreaker method for encoding triangle meshes [Rosssignac99]. Edgebreaker is one of several compression methods that encode a triangle mesh by constructing a triangle-spanning tree, and by labeling each triangle with a label that encodes both the structure of the tree and the information necessary to reconnect the triangles to recover the original mesh.

The particular labels that Edgebreaker assigns to a triangle depends on which of the adjacent triangles and bounding vertices have been previously visited. The algorithm enters each triangle from a "gate edge." The gate edge and its vertices have been previously visited, so labels only encode the status of the vertex opposite the gate and the triangles to the left and the right. A triangle has five possible combinations of previously visited and unvisited edge and vertices, represented using the following five labels:

- C – a triangle introducing a previously unvisited vertex and two unvisited neighbors
- R – a triangle whose third vertex and right neighbor have been previously visited
- L – a triangle whose third vertex and left neighbor have been previously visited
- S – a triangle whose third vertex has been previously visited but which has two unvisited neighbors. Such a triangle is a split in the triangle spanning tree
- E – a triangle whose neighbors and vertices have all been previously visited. An E triangle is a leaf in the spanning tree.

**Figure 1**: **preconditions and postconditions for each of Edgebreaker's CLERS labels.**

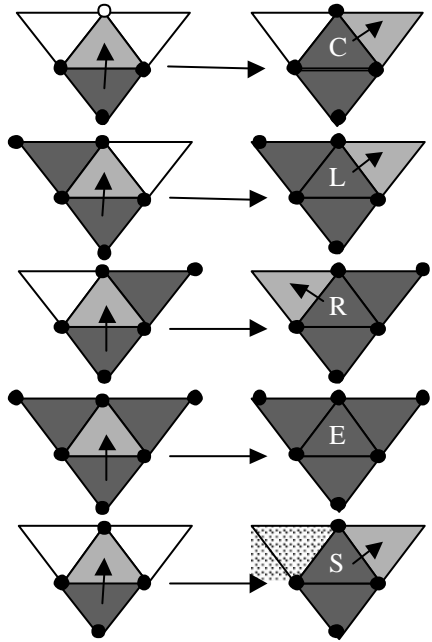

After each C, L, S, or R triangle has been processed, Edgebreaker chooses one of the triangle's neighbors as the next current triangle. For a C, L, or S, it uses a right-turn rule, moving to its right neighbor. For an R, the right neighbor has been previously visited, so Edgebreaker chooses the left neighbor instead. For an E, both neighbors have been visited, so the algorithm returns to the left neighbor of the preceding S triangle, in order to continue the encoding from another branch in the triangle spanning tree.

In Figure 1, the left column shows the precondition for each one of the 5 Edgebreaker operations, and the right column shows the result of each operation and the label associated with the most recently visited triangle. The figure uses the following shading codes. The current triangle is grey. Previously processed triangles are black. References to dark grey triangles are pushed on the stack and will be used to identify the current triangle after each E operation. White triangles have not been processed. Solid circles identify labeled vertices, and an open circle identifies a vertex that has not previously been labeled. We use incremental labels, which define the order in which vertex coordinates and other attributes are encoded. The resulting CLERS sequence suffices to capture the connectivity of the mesh (i.e. of the associated planar triangle graph).

## *Extension to Quads*

Generalizing the algorithm to quads (or other polygons) simply requires defining a more complex set of label to represent which edges and vertices have been previously visited. For a triangle, there are five possible combinations of visited and not-yet-visited edges and vertices, corresponding to the five labels above (the gate edge is always previously visited). For other polygons, the number of combinations may be computed via a recurrence relation based on the fact that no unvisited vertex may be incident to a previously visited edge. The solution to this recurrence for a polygon with n edges is the Fibonacci number F(2n-1). For a quadrilateral, therefore, there are F(7)=13 possible combinations of visited and not-yet-visited edges and vertices.

The number of combinations may be verified as follows. Mark the right edge of an n-edge polygon 0 if it its previously unvisited and 1 if previously visited. Mark each other edge with a 01 if both the edge and its right vertex are unvisited, a 10 if both the edge and its right vertex are previously visited, and a 00 if the edge is new but its right vertex has been visited. Since no visited edge follows an unvisited vertex, the digits form a 2n-1 digit number in which no two 1's are adjacent. It is known that the number of such 2n-1 digit number with no adjacent 1's equals the Fibonacci number F(2n-1).

To label the 13 possible combinations of visited edges and vertices in each quad, we split the quad into two triangles according to a splitting rule guaranteeing that the triangles are adjacent in the resulting triangle-spanning tree. We then label each quad with the two CLERS labels of the triangles resulting from the split. Since the Edgebreaker traversal always turns to the right when possible, the constraint that the two triangles remain adjacent in the spanning tree may be satisfied by splitting each quad along a diagonal so that its second triangle in the ordering is to the right of its first triangle. The split is accomplished as shown in Figure 2.

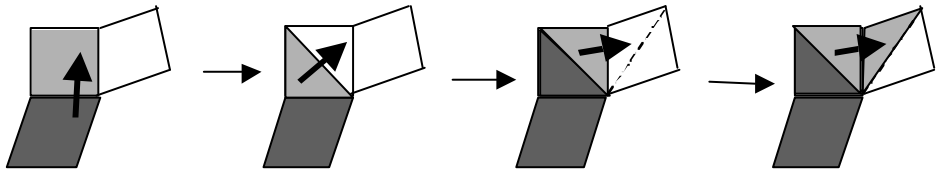

**Figure 2 Each quad is split so that the second half of the quad becomes the right neighbor of the first half. Since the second half has not been previously visited, the first half cannot be an R or an E**

Since the edge added by the split is new and by definition not previously visited, the splitting rule above ensures that the first label in a quad can never be R or E. Furthermore, it ensures that as in a triangle mesh, the second label cannot be L or E if the first label is C (Figure 3). The splitting rule therefore leaves only 13 possible label pairs: CR, CC, LE, CS, SC, LC, SE, LL, LR, LS, SL, SR, and SS. For comparison, consider that there are 23 possible pairs of adjacent symbols in the Edgebreaker code for a triangle mesh, only 18 of which correspond to adjacent triangles in the mesh, since any pair starting with E includes two triangles from different branches of the spanning tree.

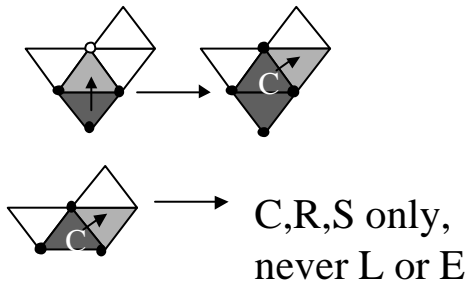

**Figure 3 Since a newly introduced vertex may not have any unsisited edges, C may never be followed by L or E**

C,R,S only, never L or E

Figure 4 shows the compression process traversing a small region of a quad mesh and splitting the resulting quads. The growing label strings are displayed adjacent to the corresponding compression stages. Further details of the compression process are given in [Rossignac99].

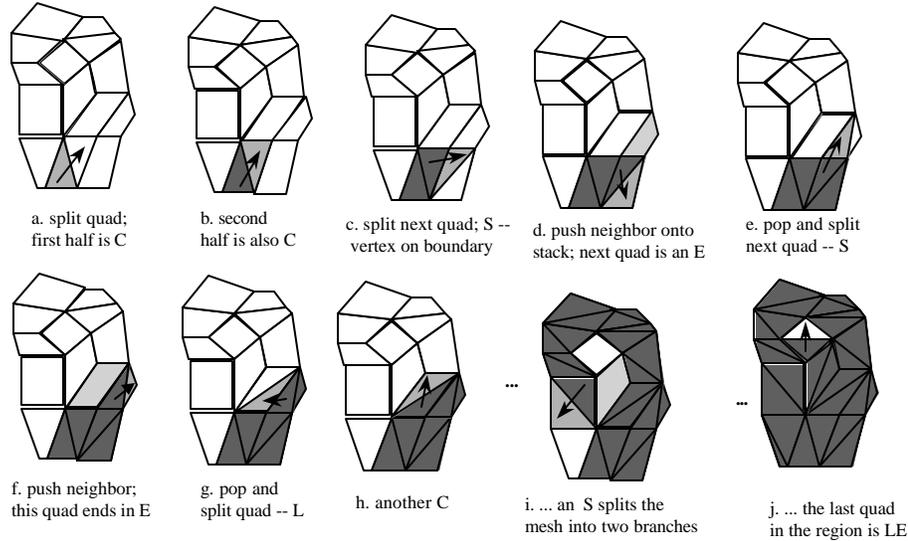

a. split quad; first half is C

b. second half is also C

c. split next quad; S -- vertex on boundary

d. push neighbor onto stack; next quad is an E

e. pop and split next quad -- S

f. push neighbor; this quad ends in E

g. pop and split quad -- L

h. another C

i. ... an S splits the mesh into two branches

j. ... the last quad in the region is LE

Figure 4 -- Encoding a Q-mesh. The region pictured is a portion of a larger Q-mesh. The current quad is light gray; previously visited quads are either dark gray or not pictured; not-yet-visited quads are white; and quads pushed onto the stack after each S have a diagonal pattern. The resulting CLERS string for this region is: CCSESELCCRCRSESECRCRSRLESELE.

## *Label Frequencies*

The definitions of the CLERS labels imply several constraints on the label frequencies, which may be used to analyze Edgebreaker's performance and to develop compact binary representations for the CLERS string [Rossignac99] [Rossignac&Szymczak99] [King&Rossignac99]. We use $|C|,|L|,|E|,|R|,|S|$ to denote the number of times each label appears in an encoded mesh, and we use $|CC|,|CS|$, etc. to denote the number of times the corresponding quad appears.

First, note that except for the two vertices adjacent to the initial gate edge, only a C label may introduce a new vertex (Figure 5). Since each vertex must be introduced before the mesh traversal is complete, and since Equation (2) implies that Q=V-2:

$|C| = V-2 = Q$ (4)

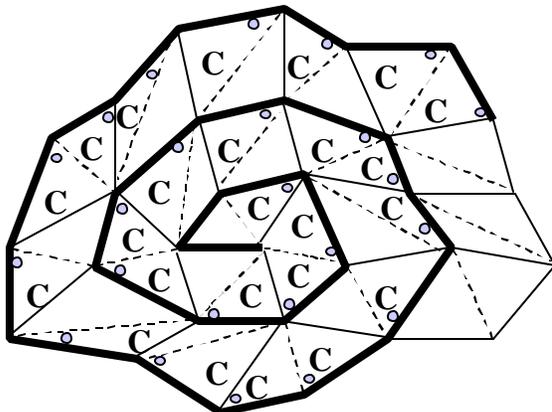
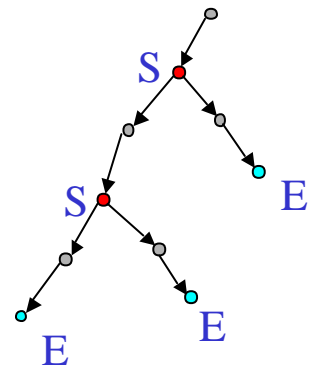

Figure 5: Each C may be associated with a unique vertex, indicated by a light gray dot adjacent to the vertex and in the C triangle.

Figure 6: In the polygon spanning tree, each S introduces a new branch and each E closes a branch. There are therefore S+1 E's.

*King, Rossignac,& Szymczak, "Connectivity Compression for Irregular Quad Meshes," GVU Tech Report GVU-GIT-99-36.*Second, each branch in the polygon spanning tree ends in an E and each branch begins either at the root or at an S (Figure 6). Therefore:
$|S| = |E| - 1$ (5)

Finally, no quad beginning with L (LC,LS,LR,LL,LE) may follow either a CR quad or a quad ending in C (CC,SC,LC), unless the two quads share a valence-two vertex that is not on a boundary of the mesh. In practice, such internal valence-two vertices are extremely rare, and most algorithms for generating Q-meshes never create them.

## An Encoding with a Worst-case bound of 3 bits/vertex

After computing the CLERS label string for a quad mesh, our algorithm encodes the labels efficiently as a binary string. Although Huffman or arithmetic coding may encode the strings of some large meshes more efficiently, we present here a single variable-length code that is guaranteed to encode the labels for ANY simple Q-mesh without internal valence-two vertices in less than 3 bits per vertex. Our code uses 22 bit patterns, organized into two tables, one for quads beginning with L and one for quads beginning with C or S. Each bit pattern identifies both the two labels of the current quad and whether the next quad begins with L or whether it begins with C or S. The decoder uses this information on how the next quad begins to determine which table to use to interpret the next bit pattern in the sequence.

**Table 1**

| Encoding | | Encoding A | |
|---|---|---|---|
| Current Quad labels | Next Quad begins with | Code | # bits |
| CC | C or S | 0 | 1 |
| CR | C or S | 100 | 3 |
| SE | C or S | 1010 | 4 |
| SE | L | 1011 | 4 |
| CS | C or S | 1100 | 4 |
| CS | L | 1101 | 4 |
| SC | C or S | 11100 | 5 |
| SS | C or S | 111010 | 6 |
| SS | L | 111011 | 6 |
| SL | C or S | 111100 | 6 |
| SL | L | 111101 | 6 |
| SR | C or S | 111110 | 6 |
| SR | L | 111111 | 6 |
| LE | C or S | 00 | 2 |
| LE | L | 01 | 2 |
| LR | C or S | 1000 | 4 |
| LR | L | 1001 | 4 |
| LS | C or S | 1010 | 4 |
| LS | L | 1011 | 4 |
| LL | C or S | 1100 | 4 |
| LL | L | 1101 | 4 |
| LC | C or S | 111 | 3 |

The code achieves its efficiency by exploiting the properties of the CLERS sequence in two different ways. First, its two-table structure allows it to take advantage of the fact that no quad beginning with L may follow a quad ending with C. Since no L follows a C, it needs only one bit pattern for each label pair ending in C, a savings over the two bit patterns needed for each label pair ending in E, R, S, and L. Second, the lengths of its bit patterns are carefully chosen so that label pairs that may be proven to appear frequently use two or three bits and so that only label pairs that may be proven to appear less frequently use more than three bits.

We use an amortization analysis [Cormen90] to prove the worst-case cost of 3 bits per vertex, by showing that the cost of each bit pattern using more than 3 bits is balanced by an offsetting number of bit patterns requiring only 2 bits. We construct the proof by arranging the quads into groups in such a way that the total cost of encoding each group is less than or equal to 3 times the number of quads in the group. We use two constraints (Equation 4 and Equation 5) on the label frequencies to group the quads and to prove that every quad in the mesh may be included in one of these groups. Since all but one of the quads in the mesh belongs to a group with an average cost between 2 bits/quad and 3 bits/quad, the total encoding cost is guaranteed to be between 2V and 3V bits.

The first grouping step uses the constraint that the total number of C labels equals the total number of quads (Equation 4). Since each quad contains either 0, 1, or 2 labels of type C and since there are exactly Q such labels altogether, there must be an equal number of quads containing two C's and of quads containing 0 C's. We may therefore pair each quad containing no C's with a CC quad. For example, we group each LL with a CC to obtain a group (LL,CC); we may group each LE with a CC to obtain (LE,CC), and we may group each SS to obtain (SS,CC).

Note that the first grouping step is applied to every quad containing no C's and to every CC quad. The only quads that are not affected by the first step are those containing exactly one C. Two of the pairs containing exactly one C, namely CR and LC, are encoded in three bits or less and do not need to be grouped.

This first step is sufficient for three label pairs: LL, SE, and LR. Each LL, SE, or LR quad is encoded in four bits, and each CC is encoded in one bit. The total cost of encoding an (SE, CC), (LR,CC) or (LL,CC) grouping is therefore 4+1=5 bits for two quads, for an amortized cost of 2.5 bits per quad.

This first step is not sufficient, however, for any of the quads containing one or more S's, except for SE. For example, the grouping (SS,CC) has a total cost of seven bits, which is more than three bits per quad. Furthermore, the pairs CS and SC cost four and five bits each, and they are not affected by the first grouping step.

We address the cost of the remaining labels, therefore, with a second grouping step based on Equation 5 above. Since every S begins a new branch and since only an E may terminate a branch, every S must have exactly one offsetting E. We therefore pair each S that is not part of an SE quad with the LE quad that terminates the branch that began at the S.

Note that since the second grouping step is performed after the first step, it is applied to the results of the first step. For example, since the first rule groups every LE with a CC, applying the second grouping step to SC and CS produces the groupings (SC,LE,CC) and (CS,LE,CC), not (SC,LE) or (CS,LE). Since the LE costs 2 bits, the CC costs 1 bit, and the CS costs 4 bits, the grouping (CS,LE,CC) costs 4+2+1=7 bits for an amortized cost of 7/3 bits per quad. Likewise, the grouping (SC,LE,CC) costs 5+2+1=8 bits, for an amortized cost of 8/3 bits per quad.

Likewise, the second rule groups an (SL,CC) with an (LE,CC) to obtain the combined grouping (SL,CC,LE,CC), with a total cost of 6+1+2+1=10 bits for 4 quads, or 2.5 amortized bits/quad.

Applying these constraints to each of the 13 possible label pairs, we obtain the following groupings and amortized costs:

**Table 2**

| Grouping | Total Cost | # Quads in group | Amortized Cost | # bits saved | # occurrences of this group |
|---|---|---|---|---|---|
| (LL,CC) or (SE,CC) or (LR,CC) | 5 | 2 | 2.5 | 1 | |LL|+|SE|+|LR| |
| (SC,LE,CC) | 8 | 3 | 2.67 | 1 | |SC| |
| (CS,LE,CC) | 7 | 3 | 2.33 | 2 | |CS| |
| (SL,LE,CC,CC) or (SR,LE,CC,CC) | 10 | 4 | 2.5 | 2 | |SL|+|SR| |
| (LS,LE,CC,CC) | 10 | 4 | 2.5 | 2 | |LS| |
| (SS,LE,LE,CC,CC,CC) | 13 | 6 | 2.17 | 5 | |SS| |
| (LC) | 3 | 1 | 3 | 0 | |LC| |
| (LE,CC) – the final LE has no matching S | 3 | 2 | 1.5 | 3 | 1 |
| (CR) | 3 | 1 | 3 | 0 | |CR| |

The total cost of encoding a Q-mesh with the above code may be found by summing the amortized costs of each quad. The amortized cost is exactly three bits for the quads in only two of the groupings in Table 2. Meanwhile, each of the other seven groupings has a total cost at least one bit below three bits per quad: one bit is saved for each LL, SE, LR, and SC, two bits are saved for each CS, SL, LS, and SR, three bits are saved for the final LE, and five bits are saved for each SS. The total cost of the encoding, therefore, equals 3Q -- |LL| -- |SE| -- |LR| -- |SC| -- 2|CS| -- 2|SL| -- 2|SR| -- 2|LS| -5|SS| - 3.

We can simplify the above expression for the cost of the above code by noting that the savings amount to at least one bit per CC, and that |CC|= |LL| + |SE| + |LR| + |SC| + |CS| + 2 |SL| + 2 |SR| + 2 |LS| - 2 |SS| + 1. Substituting, we obtain a total cost of 3Q -- |CC| -- |CS| -- 2|SS| - 2. The value of this expression may vary, but it may never exceed 3Q. Since V=Q+2 for a quad mesh (Equation 1), the total cost is therefore guaranteed to be less than 3V bits.

### *A more complex code with a guaranteed cost of 2.67 bits/vertex*

Since the cost of the above code is bounded by 3Q -- |CC|, it clearly performs best on meshes with a large number of CC quads. In our experiments on Q-meshes below, however, we have found that over half the quads are CR, less than one fourth are CC, and less than one fourth have no C's. Consequently, one may obtain better results in practice via alternate fixed codes using fewer bits for each CR and more bits for each CC.

Furthermore, in [King&Rossignac99], King and Rossignac demonstrate that one can obtain better worst-case encoding costs for triangle meshes by using a choice of 3 different fixed codes, with two extra bits at the beginning of the encoding to signal which code to use for a given mesh. The same technique may also be applied to quadrilateral meshes to obtain better guaranteed compression ratios.

We therefore present the following three encodings for CLERS label pairs as alternatives to the above code. Choosing the best of the four fixed codes presented here guarantees a worst-case cost of 2.67 bits/vertex to encode any quadrilateral mesh with no holes, handles, or internal valence-two vertices.

The first alternative, Encoding B, switches the bit patterns for CC and CR, costing one bit per CR and three bits per CC, for a total cost of 3Q -- |CR| + |CC| -- |CS| -- 2|SS|. This alternative is usually best in practice. Encoding C makes another small change, swapping the codes for CS followed by L and for SC, giving a total cost of 3Q -- |CC| -- |SC| -- |CS followed by C or S| -- 2|SS|. Finally, encoding D makes a much larger number of changes to obtain a total cost of 3Q -- |CR| -- |LC|.

*King, Rossignac,& Szymczak, "Connectivity Compression for Irregular Quad Meshes," GVU Tech Report GVU-GIT-99-36.*To demonstrate that either Encoding A, C, or D costs less than 2.67 bits per vertex for each mesh, we sum the expressions for their costs and divide by three to compute their average cost. Clearly, the cost of the best of these three codes must always be less than or equal to their average cost.

Summing the three expressions, we find that the total cost is 9Q -- 2|CC| -- |CR| -- |LC| -- |CS| -- |SC| --|CS followed by C or S| -- 4|SS|. Now, since we know the total number of C's is equal to Q (Equation 4), 2|CC| + |CR| + |LC| + |CS| + |SC| = |C| = Q. Substituting, we find the total cost is 8Q – 4|SS| -- |CS followed by C or S|. Dividing by 3, we find:

min{cost(A),cost(C),cost(D)} <= sum {cost(A),cost(C),cost(D)}/3 <= 8Q/3 = 2.67 Q          (6)

**Table 3**

| Encoding | | Encoding B | | Encoding C | | Encoding D | |
|---|---|---|---|---|---|---|---|
| Current Quad | Next | Code | # bits | Code | # bits | Code | # bits |
| CC | C or S | 100 | 3 | 0 | 1 | 00 | 2 |
| CR | C or S | 0 | 1 | 100 | 3 | 01 | 2 |
| SE | C or S | 1010 | 4 | 1010 | 4 | 1010 | 4 |
| SE | L | 1011 | 4 | 1011 | 4 | 1011 | 4 |
| CS | C or S | 1100 | 4 | 1100 | 4 | 1100 | 4 |
| CS | L | 1101 | 4 | 11100 | 5 | 1101 | 4 |
| SC | C or S | 11100 | 5 | 1101 | 4 | 1000 | 4 |
| SS | C or S | 111010 | 6 | 111010 | 6 | 111110 | 6 |
| SS | L | 111011 | 6 | 111011 | 6 | 111111 | 6 |
| SL | C or S | 111100 | 6 | 111100 | 6 | 10010 | 5 |
| SL | L | 111101 | 6 | 111101 | 6 | 10011 | 5 |
| SR | C or S | 111110 | 6 | 111110 | 6 | 11100 | 5 |
| SR | L | 111111 | 6 | 111111 | 6 | 11101 | 5 |
| LE | C or S | 00 | 2 | 00 | 2 | 000 | 3 |
| LE | L | 01 | 2 | 01 | 2 | 111 | 3 |
| LR | C or S | 1000 | 4 | 1000 | 4 | 1000 | 4 |
| LR | L | 1001 | 4 | 1001 | 4 | 1001 | 4 |
| LS | C or S | 1010 | 4 | 1010 | 4 | 001 | 3 |
| LS | L | 1011 | 4 | 1011 | 4 | 101 | 3 |
| LL | C or S | 1100 | 4 | 1100 | 4 | 1100 | 4 |
| LL | L | 1101 | 4 | 1101 | 4 | 1101 | 4 |
| LC | C or S | 111 | 3 | 111 | 3 | 01 | 2 |

**Table 4 – Analysis of Encoding D**

| Groupings in Encoding D | Total | # Quads | Amortized | # occurrences of this group |
|---|---|---|---|---|
| (LL,CC) or (SE,CC) or (LR,CC) | 6 | 2 | 3 | |LL|+|SE|+|LR| |
| (SC,LE,CC) | 9 | 3 | 3 | |SC| |
| (CS,LE,CC) | 9 | 3 | 3 | |CS| |
| (SL,LE,CC,CC) or (SR,LE,CC,CC) | 12 | 4 | 3 | |SL|+|SR| |
| (LS,LE,CC,CC) | 10 | 4 | 2.5 | |LS| |
| (SS,LE,LE,CC,CC,CC) | 18 | 3 | 3 | |SS| |
| (LC) | 2 | 1 | 2 | |LC| |
| (LE,CC) – the final LE has no matching S | 5 | 2 | 2.5 | 1 |
| (CR) | 2 | 1 | 2 | |CR| |

# Decompression

Wrap&Zip decompression [Rossignac&Szymczak99] decodes and uses the labels of the CLERS string to decide where to append each new triangle to a previously reconstructed one. The result is a simply connected topological polygon that corresponds to a triangle-spanning tree of the original mesh. To correctly glue the corresponding pairs of its bounding edges, Wrap&Zip uses the CLERS labels to orient the free edges that bound the polygon clockwise for L, R, and E, and counter-clockwise for C triangles, as shown in Figure 7.

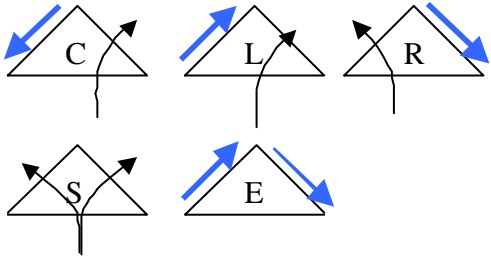

**Figure 7 Wrap&Zip decompression assigns an orientation to each of the edges marked by a straight arrow, and then it glues together edges whose arrows point away from the same vertex**

A recursive procedure restores the complete incidence information by gluing pairs of adjacent edges whose orientations point away from their common vertex. Vertices are labeled in the order in which they are first encountered and decoded accordingly. Since the splitting rule ensures that the second triangle in each quad appears immediately after the first triangle in the encoded string of CLERS symbols, the original quads may be restored during decompression by merging each pair of consecutive triangles into a quad.

# Extension to meshes with holes, handles, and triangles

The 3 bits per vertex encoding and the above descriptions of the compression and decompression algorithms guarantee good worst-case performance on a simple Qmesh, with no holes, handles, or triangular faces. In practice, however, predominantly quadrilateral meshes may have triangular faces, holes, and handles. It is therefore important to extend our approach to meshes of complex topology and to QT-meshes.

## *Holes*

A hole is a simply connected loop of boundary edges (edges incident to only one face). For the experiments reported below, we use the method introduced in [Rossignac99]. This approach uses an S label to represent the first triangle (or half of a quad) that introduces a vertex incident to a hole. After labeling that S, the compression algorithm marks all the other vertices and edges in the hole as visited, thus merging the bounding loop of the hole into the loop bounding the union of all not-yet-visited faces. A table indicating which S's introduce holes and indicating the number of vertices in each hole is transmitted at the beginning of the compressed bitstream.

We have also explored a second alternative based on the method proposed by Touma and Gotsman to encode the holes [Touma&Gotsman98]. We create a dummy vertex for each hole and patch the hole by adding a triangle connecting each boundary edge to the dummy vertex. (For holes with even numbers of edges, we join the new triangles into quads). A list of the dummy vertex identifiers may be transmitted at the beginning of the compressed bitstream, and the faces incident to the dummy vertices may be deleted after decompression.

## *Handles and Holes*

For meshes with both holes and handles, we use the method originally described in [Rossignac99] and improved in [Rossignac&Szymczak99]. In a mesh with handles, a face containing as S label as defined above may not separate the union of all not-yet-visited faces into multiple connected components. Instead, an S face may create a new hole (a new connected component of the boundary edges). For each such hole-creating S, there is another S that merges two bounding loops into one. Both the hole-creating and the hole-merging S's must be distinguished from regular S's and must be associated with an integer that identifies which of the vertices on the bounding loop they meet.

## QT-meshes

Since the Q-mesh compression algorithm above uses the same CLERS labels Edgebreaker uses for triangles, it may be generalized to meshes containing both quads and triangles by using a single CLERS label for each triangle and using one of the 13 CLERS label pairs above for each quad. For applications which do not require preserving the original quads (such as transmission for remote display), one may use the compression algorithm above without modification to construct the CLERS string for a QT-mesh. One may use the standard Wrap&Zip algorithm to decompress the string, without joining any triangles into quads. The worst-case cost of this approach is the same as the worst-case cost of encoding a triangle mesh using Edgebreaker: 3.67 bits/vertex [King&Rossignac99].

Meanwhile, for applications in which it is necessary to preserve the original connectivity of a QT-mesh, the encoding must be modified to store additional information to indicate which labels belong to triangles and which should be joined to form quads. Several methods are available to modify the encoding. For example, one bit per face may be used to indicate whether the face is a triangle or a quad, and these polygon-size bits may be entropy coded separately from the label string.

For many QT-meshes encountered in practice, we have found that the best alternative is to use a sixth label, T, to identify a triangle face, thus representing each triangle as a label pair TC, TL, TE, TR, or TS. By ensuring that each label pair represents a single face, this approach simplifies the design of the decoder and it makes it easy to exploit the relationships between adjacent labels and adjacent quads. In practice, many QT-meshes have triangles as only a small fraction of their faces, with 8-12% triangles common in finite-element meshes. This method is most effective for such predominantly quadrilateral meshes in which the cost of the extra T labels is low. We use this approach in the experiments reported below.

## Valence-Two Vertices

As noted above, our worst-case bound of 2.67 bits per vertex only strictly applies for quad meshes without internal valence-two vertices. Although valence-two vertices are rare in practice and undesirable in many applications, one cannot guarantee that a quad mesh encoder will never encounter one. We provide, therefore, a simple modification to our algorithm that guarantees to encode Q-meshes with valence-two vertices in 3.07 bits per vertex or less.

The modification is based on noting that every valence-two vertex has exactly two incident quads, each of which has only two edges that are not incident to the valence-two vertex. By deleting the valence-two vertex and its incident edges, therefore, one may obtain a new Q-mesh in which the two quads incident to the deleted vertex have been replaced by their union, a single quadrilateral. Furthermore, one may recover the original mesh by reversing this step, as long as one knows which face to split and which of its vertices should share edges with the reinserted vertex.

To encode Q-meshes with valence-two vertices, we apply this removal step iteratively until we obtain a modified mesh with no valence-two vertices. We then encode both the modified mesh (using the 2.67 bit code above) and a table of auxiliary information indicating which faces to split and how to connect their vertices to reinserted vertices to recover the original mesh.

Let Q be the number of quads in the original mesh, V be the total number of vertices, and V' be the number of valence-two vertices. The modified mesh contains V-V' vertices and may therefore be encoded in 2.75(V-V') bits. The auxiliary table, meanwhile, may use one bit per quad to indicate whether or not it should be split and one bit per split quad to indicate which pair of its vertices should share edges with the reinserted vertex, for an additional cost of V' + Q bits. The auxiliary table may be further compressed by entropy coding the string of bits indicating which quads to split.

With entropy coding, the auxiliary table requires V' - V' log(V'/Q) + (Q-V') log(1-V'/Q) bits, for a total cost of 2.67 V – 1.75 V' - V' log(V'/Q) + (V'-Q) log(1-V'/Q) bits. Using a numerical solver, one may easily calculate that the maximum of this expression is less than 3.07 V bits, with the maximum occurring near V'/Q=0.23. We note that while this result is not as impressive as our 2.67 V result for Q-meshes without internal valence-two vertices, it is still lower than the theoretical limit of 3.24 V bits for the worst-case cost of encoding a triangle mesh.

### *Polygon Meshes*

Our algorithm may also be extended to polygon meshes via a generalization of our quad splitting rule. Just as a quad may be triangulated with a split that keeps the two triangles adjacent in the spanning tree, we split each polygon into a fan of triangles, with splits that ensure each pair of triangles adjacent in the polygon are adjacent in the spanning tree. We then label each triangle with one of the five CLERS labels. As in the quadrilateral case, only the final triangle of a polygon may be R or E, and, for meshes with no valence-two vertices, no L or E may follow a C.

We record the size of each polygon by marking each C, S and L triangle with an extra bit indicating whether to join that triangle with the next triangle in the sequence to form a larger polygon. R and E triangles do not require an extra bit, since they may only occur as the last label in a face. One of the codes presented in [King&Rossignac99], meanwhile, may be used to encode the CLERS labels with a constant cost of 4V – |S| – |L| – 2. Since |C|=V– 2 by Equation 4, the total cost of the resulting encoding is (V – 2 + |S| + |L|) + (4V – |S| – |L| – 2), or less than 5 bits per vertex.

This guaranteed performance of 5 bits per vertex or less applies to any polygon mesh without internal valence-two vertices, or, equivalently, to any triconnected polygon mesh. Furthermore, since the dual graph of any triconnected mesh is also a triconnected mesh, we can achieve further improvements by using He and Kao's method of using one bit to choose between encoding the original mesh and encoding its dual [He&Kao99], for a cost of 5 min{V,F} bits or no more than 2.5 bits per edge.

For polygon meshes with valence-two vertices, an L or an E may follow a C if and only if that C is the last label in a polygon and it introduces an internal valence-two vertex.

## **Experimental Results for Meshes Encountered in Practice**

In addition to the topological complexities described above, meshes encountered in practice are often highly regular, with many valence-four vertices. It is therefore often more important in practice to obtain high compression ratios by exploiting regularities than to obtain good worst-case results on highly irregular meshes. Furthermore, some authors using entropy coding on triangle meshes have reported better compression ratios than the 2.67 bits per vertex worst-case guarantee our fixed code provides for Q-meshes. To evaluate our algorithm, it is thus necessary both to determine its effectiveness on highly regular meshes, and to compare its performance in practice to an alternative approach based on randomly splitting quads into triangles.

In this section, therefore, we present results comparing our approach to an alternative based on random splits on a variety of quad meshes used in practice:

1. Q-meshes constructed via Nowottny's algorithm for finite-element mesh generation [Nowottny99].
2. Irregular Q-meshes constructed from a T-mesh by adding a vertex at the center of every triangle and in the center of every edge to subdivide each triangle into three quads. Some authors have used this method to create Q-meshes from triangle mesh models for finite-element applications [Mueller-Hannemann99].
3. QT-meshes from Viewpoint's 3D model library [Viewpoint].

4. Highly regular Q-meshes generated via parametric functions or via Catmull-Clark subdivision [Catmull&Clark78].

**Figure 8 – Some wireframe closeups of the sample meshes**

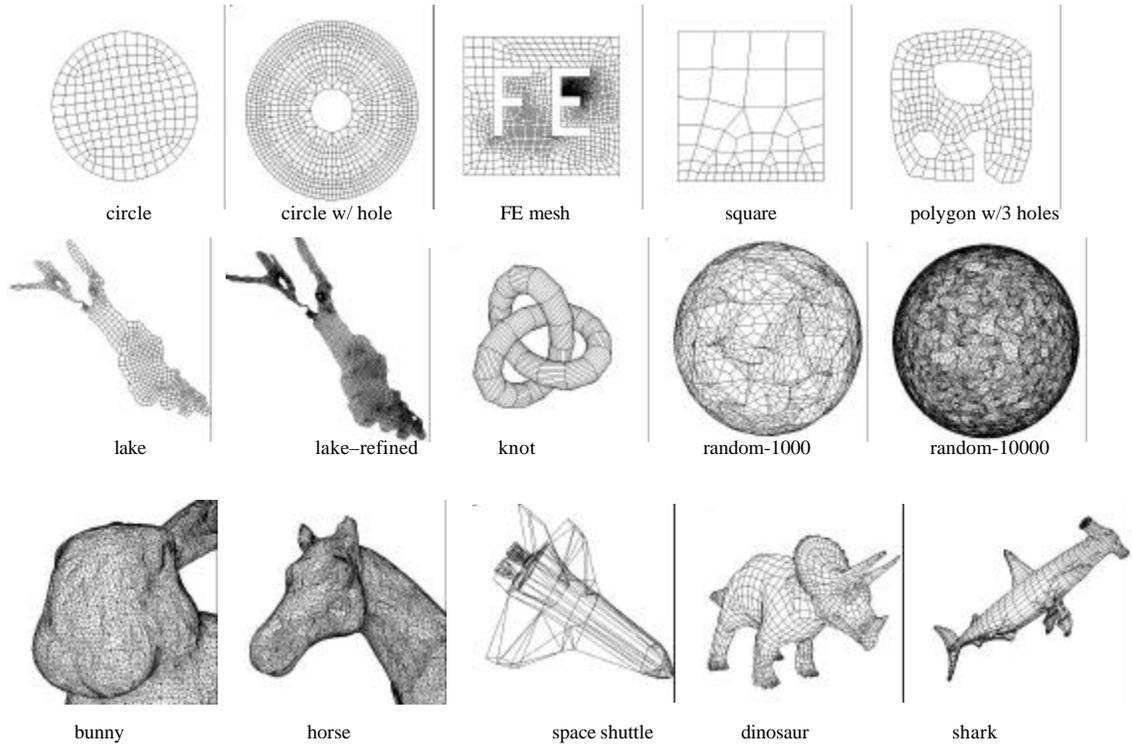

Figure 8: Quad meshes used in testing. From top left to bottom right: circle, circle w/ hole, FE, square, polygon w/ 3 holes, lake constance, lake constance/refined, knot, mesh from convex hull of 1000 points, mesh from convex hull of 10,000 points, closeup of bunny, closeup of horse, QT-mesh of space shuttle, QT-mesh of triceratops, QT-mesh of shark

The meshes of Type 2 (subdivided triangles) that we use include two models, a bunny and a horse, and several semi-random meshes created by taking the convex hull of points randomly distributed on the unit sphere. The semi-random meshes of Type 2 are the most irregular meshes we use in our tests, but contain a high proportion of valence-four and valence-three vertices.

Some of the original Viewpoint models of Type 3 contain a few polygons in addition to triangles and quads. To create a valid QT-mesh, we split each polygon with more than 4 edges into one or more quads and zero or one triangles.

For the fixed code in the experiments below, we use Encoding B above, which typically has the best performance in practice.

### *Entropy coding*

Several techniques such as Huffman coding and arithmetic coding are available to obtain better compression ratios by exploiting the frequencies of the symbols in a datastream. These algorithms may achieve efficiencies arbitrarily close to but not less than the entropy of the datastream [Press93]. The entropy may be computed as

$$\text{entropy} = \Sigma(-P(x=i)\log(P(x=i))) \qquad (7)$$

where x the next label in the string, i is one of the CLERS labels, and P(x=i) is the ratio of the number of times the ith label appears to the total length of the string.

We use a variation on arithmetic coding in which both the encoder and the decoder have a memory of one or more labels immediately preceding the current label. The encoder and decoder may therefore use different bit codes for the same label, depending on the contents of the memory, in order to take advantage of patterns in the CLERS string. With this approach, the entropy of the CLERS string (which determines the compression ratio) is a function of the conditional probability of each label given the identity of the previous labels. It may be computed as

$$\text{entropy} = \Sigma(-P(x=i|y_1=j_1,..y_n=j_n)\log(P(x=i|y_1=j_1,..y_n=j_n))) \qquad (8)$$

where x is the next label in the string, $y_1…y_n$ are the n labels preceding x in the CLERS string, $j_1…j_n$ are each equal to one of the CLERS labels, and $P(x=i|y_1=j_1,..y_n=j_n)$ is the conditional probability of the next label being i given that $y_1=j_1,..y_n=j_n$. The summation is performed over all possible combinations of i and $j_1…j_n$.

## *Table of Results*

We first compare the performance of our fixed code to an approach based on randomly splitting quads into triangles. The goal of these experiments is to determine whether an approach that does not preserve the original quads can use entropy coding to achieve compression ratios similar to those that our fixed code achieves. We implement the random split approach by randomly triangulating each quad and applying Edgebreaker to the resulting triangle mesh.

Table 5 presents the results. For the quad-preserving approach, it reports the actual cost of encoding the mesh via the fixed code that does not allow valence-two vertices. For the random split approach, it reports the entropy values computed via Equation 7 for the triangle mesh's CLERS string. Note that these entropy values are lower bounds on the efficiency of entropy coding, and they do not include the cost of representing a codeword dictionary or frequency table. The actual cost of encoding the mesh via random splits and entropy coding may therefore be higher than the entropy values presented here, but it may not be lower.

**Table 5**

| Model Name | Type | vertices (V) | quads (Q) | holes (H) | Fixed Code (bits/V) | Random Splits – 1 label memory | Random Splits – 3 label memory |
|---|---|---|---|---|---|---|---|
| circle | 1 | 202 | 180 | 1 | 1.58 | 2.20 | 1.87 |
| circle w/hole | 1 | 1002 | 928 | 2 | 1.29 | 2.16 | 1.70 |
| FE mesh | 1 | 3153 | 2996 | 3 | 1.55 | 2.35 | 2.00 |
| square | 1 | 85 | 66 | 1 | 1.90 | 2.67 | 1.33 |
| polygon w/ holes | 1 | 270 | 220 | 4 | 1.74 | 2.63 | 1.71 |
| lake | 1 | 909 | 765 | 2 | 1.81 | 2.87 | 2.02 |
| lake-refined | 1 | 7304 | 6876 | 2 | 1.36 | 2.32 | 1.94 |
| knot | 4 | 900 | 900 | 0 | 1.32 | 2.24 | 1.82 |
| random-1000 | 2 | 5990 | 5988 | 0 | 2.02 | 2.86 | 2.29 |
| random-5000 | 2 | 29990 | 29988 | 0 | 2.01 | 2.86 | 2.32 |
| random-10000 | 2 | 59990 | 59988 | 0 | 2.01 | 2.86 | 2.31 |
| bunny | 2 | 90962 | 90960 | 0 | 1.98 | 2.82 | 2.30 |
| horse | 2 | 62240 | 62238 | 0 | 1.96 | 2.79 | 2.30 |

The results show that our quad-preserving approach clearly outperforms random splits. Our fixed code consistently achieves 28-40% smaller encodings than the minimum cost random splits may achieve with a 1-label memory. Using a 3-label memory for entropy coding, the entropy of using random splits is lower than the fixed code cost for only one of the sample meshes. The highest costs of both approaches occur on meshes of type 2, with our fixed code requiring up to 2.02 bits per vertex, and with random splits having an entropy value up to 2.31 bits per vertex.

We next consider the efficiency of using entropy coding with our quad-preserving approach, in order to see if our approach can achieve even better results via entropy coding. In Table 6, we compare the entropy values measured for our quad-preserving approach to those measured for the random-split approach. Note that these entropy values do not include either the overhead of representing the codeword dictionaries or the cost of indicating handles and holes, both of which are the same or nearly the same for both approaches.

For all but the most regular meshes (type 4), the results we achieve with entropy coding and a 1-label memory are a slight improvement at best over our fixed code. With a 3-label memory, however, the quad-preserving method outperforms random splits even more dramatically. Its entropy values range from 0.24 to 0.85 bits per vertex, a savings of 60% to 85% over random splits for all but one sample mesh.

**Table 6**

| Model Name | Type | vertices (V) | quads (Q) | tris (T) | holes (H) | Quad Method – 1 label memory | Quad Method – 3 label memory | Random Splits – 1 label memory | Random Splits – 3 label memory |
|---|---|---|---|---|---|---|---|---|---|
| circle | 1 | 202 | 180 | 0 | 1 | 1.63 | 0.38 | 2.20 | 1.87 |
| circle w/hole | 1 | 1002 | 928 | 0 | 1 | 1.31 | 0.24 | 2.16 | 1.70 |
| FE mesh | 1 | 3153 | 2996 | 0 | 3 | 1.67 | 0.51 | 2.35 | 2.00 |
| square | 1 | 85 | 66 | 0 | 1 | 2.39 | 0.85 | 2.67 | 1.33 |
| polygon w/3 holes | 1 | 270 | 220 | 0 | 4 | 2.16 | 0.64 | 2.63 | 1.71 |
| lake | 1 | 909 | 765 | 0 | 2 | 2.33 | 0.71 | 2.87 | 2.02 |
| lake-refined | 1 | 7304 | 6876 | 0 | 2 | 1.36 | 0.30 | 2.32 | 1.94 |
| knot | 4 | 900 | 900 | 0 | 0 | 0.92 | 0.35 | 2.24 | 1.82 |
| random-1000 | 2 | 5990 | 5988 | 0 | 0 | 1.98 | 0.79 | 2.86 | 2.29 |
| random-5000 | 2 | 29990 | 29988 | 0 | 0 | 1.94 | 0.77 | 2.86 | 2.32 |
| random-10000 | 2 | 59990 | 59988 | 0 | 0 | 1.90 | 0.74 | 2.86 | 2.31 |
| bunny | 2 | 90962 | 90960 | 0 | 0 | 1.87 | 0.73 | 2.82 | 2.30 |
| horse | 2 | 62240 | 62238 | 0 | 0 | 1.84 | 0.72 | 2.79 | 2.30 |
| space shuttle | 3 | 310 | 223 | 170 | 0 | 4.14 | 1.14 | 2.54 | 1.90 |
| dinosaur | 3 | 2832 | 2266 | 346 | 0 | 3.00 | 1.03 | 2.58 | 2.15 |
| shark | 3 | 2668 | 2280 | 316 | 0 | 2.53 | 0.78 | 2.39 | 2.04 |

For QT-meshes, our approach needs a 3-label memory to outperform random splits. With only a 1-label memory, the cost of marking the original quads and triangles exceeds the savings of splitting quads via a regular rule. A 3-label memory, however, allows our method to exploit patterns among the quad label pairs even when some quads are separated by triangles. With a 3-label memory, our approach outperforms random splits on QT-meshes by 40% to 62%, with entropy values from 0.78 to 1.14 bits per vertex.

We have performed additional tests on meshes produced by subdividing the meshes in the table above. We have found that with entropy coding, the efficiency of our method increases by a factor of sqrt(2) for each level of Catmull-Clark subdivision applied to a mesh.

## Comparison to Information-theoretic lower bounds

Our results on Q-meshes may also be evaluated by comparing them to both the worst-case costs of existing triangle mesh compression algorithms, and to the known theoretical limits on the worst-case cost of any possible triangle mesh encoding. The best worst-case cost of any known scheme for triangle meshes is 3.67 bits/vertex [King&Rossignac99]. The best possible worst-case cost of encoding a triangle mesh, meanwhile, is 3.24 bits/vertex, a consequence of Tutte's enumeration of all planar triangulations [Tutte62]. Tutte proved that the total number of triangulations with E edges is $O(2^{1.08E})$. Using Euler's formula as in Equation (3), Tutte's enumeration is equivalent to $\Theta(2^{3.24V})$. Since every compression scheme must have a unique encoding for each unique triangulation, every possible compression algorithm will require at least 3.24 bits/vertex or 1.62 bits/triangle to encode some triangle meshes [see also Keeler&Westbrook95].

Our 2.67 bits per vertex encoding, therefore, has a lower worst-case cost than any possible triangle mesh compression scheme, and it implies that there are fewer quad meshes with V vertices than triangle meshes with V vertices. It may be possible, however, that cheaper encodings of Q-meshes than ours exist. A more precise bound on the number of possible quad meshes, similar to Tutte's enumeration, is needed to assess how much our per vertex bound may be improved. While our encoding result implies that the number of quad meshes without valence-two vertices must be bounded above by $2^{2.67V}$, it does not provide a lower bound.

We obtain a lower bound on the number of Q-meshes by constructing a mapping between quad meshes and triangle meshes that allows us to compare the number of quad meshes to Tutte's result for triangle meshes. Consider the process of converting a quad mesh to a triangle mesh and its inverse. As noted above, every quad mesh may be converted into a triangle mesh by adding one edge per quad to split it into triangles. Conversely, one may convert a triangle mesh into a Q-mesh by selecting exactly one edge incident to each triangle and deleting that edge in order to merge the two triangles sharing the edge into a single quad.

We use the four-color theorem to prove that every triangle mesh may be converted into a Q-mesh by the deletion process described above. The four-color theorem, proven by Appel and more recently by Thomas [Appel89][Robertson97], states that the faces of any planar polygon mesh may be colored with four colors such that the two faces incident to each edge have different colors. A four-coloring of the dual graph of a triangle mesh, meanwhile, may be used to construct a three-color edge-coloring, in which each triangle has exactly one edge of each color [Kaufman90]. Given such a three-coloring of edges, one may produce a quad mesh by choosing one of the three colors and deleting every edge of that color. Clearly, if an edge e is deleted, the other four edges bounding the two triangles incident to e remain.

The three-color edge-coloring is constructed as follows [Kaufman90][Baez 93]. First, construct the dual graph of the triangle mesh, and compute a four-coloring of its faces, which is equivalent to a four-coloring of the vertices of the triangle mesh. Let the colors be red (R), green (G), blue (B), and white (W). Colors may be assigned to the edges via a multiplication rule, as shown in Figure 9. If one of the vertices incident to the edge is white, set the color of the edge equal to the color of the other incident vertex. If neither of the incident vertices is white, set the color of the edge equal to the non-white color which does not appear on either of the vertices (i.e. if the vertices are red and blue, the edge is green). After all the edges have been colored, each triangle will have exactly one red edge, one green edge, and one blue edge. A quad mesh may then be constructed by choosing one of the three colors and removing all the edges of that color. (Note that the quad meshes constructed in this manner may have interior valence-two vertices.)

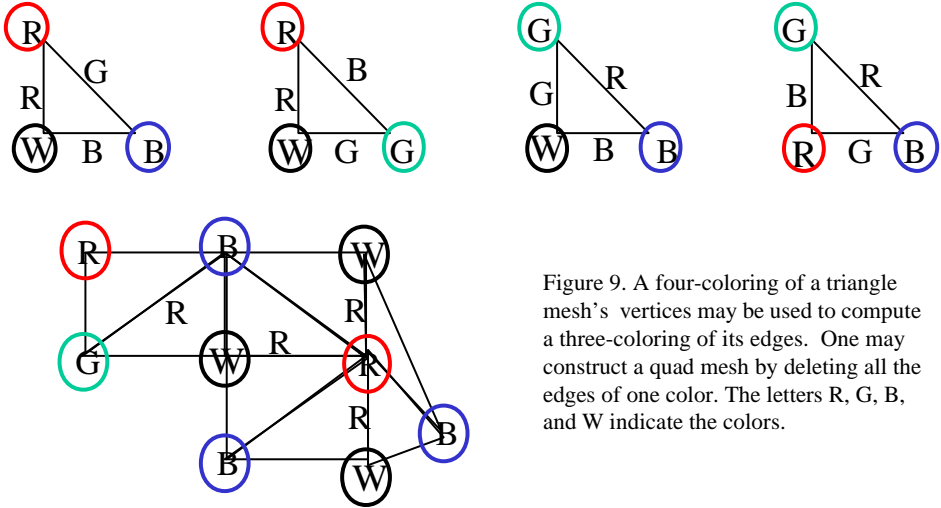

Figure 9. A four-coloring of a triangle mesh's vertices may be used to compute a three-coloring of its edges. One may construct a quad mesh by deleting all the edges of one color. The letters R, G, B, and W indicate the colors.

Every triangle mesh, therefore, may be converted into a quad mesh. One may recover the original triangle mesh by splitting each quad into triangles, using the sequence of splits that restores the deleted edges. Consequently, there is at least one quad mesh and at least one sequence of splits for every four-coloring of a triangle mesh. Since each quad mesh has $2^V$ possible splitting sequences and since each of the $2^{3.24V}$ triangle meshes has at least one four coloring, the total number of quad meshes must be at least $\Omega(2^{3.24V}/2^V)$ or $\Omega(2^{2.24V})$. As a consequence, no compression scheme may encode every quad mesh in less than $2.24 V + O(1)$ bits.

For a tri/quad mesh with T triangles and Q quads, a similar argument indicates that the number of possible tri/quad meshes with the same T and Q is at least $(2^{3.24(Q+T/2)}/2^Q)$ or $(2^{2.24Q+1.62T})$. The T/Q ratio may be encoded in $\log(V)$ bits, for a lower bound on the encoding of a tri/quad mesh of $2.24V+0.5T+\log(V)$.

## Conclusion and future work

We have presented the first compression algorithm designed specifically for Q-meshes and QT-meshes. Our algorithm has been tested on a variety of sample meshes, and it compresses Q-meshes used in practice to between 0.24 and 2 bits per vertex via entropy coding and to between 1.3 and 2 bits per vertex using a fixed code with a guaranteed worst-case cost of 2.67 bits per vertex.

Our results in both the worst-case and in entropy-coding meshes used in practice have shown that quad meshes are inherently less complex and cheaper to encode than triangle meshes with the same number of vertices. These results suggest that in applications requiring compression and level-of-detail management, it may be more efficient to support quadrilateral and polygonal meshes than to triangulate all polygons arbitrarily.

While we have only presented results on connectivity compression, it may also be possible to achieve greater compression of the vertex coordinates and other per-vertex data of Q-meshes and QT-meshes by exploiting their quadrilateral structure. Since quads created by solid modeling software are often planar or nearly planar, and since they are usually convex, their geometric properties may allow prediction rules more effective than the geometric predictors based on ancestor vertices and subdivision masks that have been used in practice [Taubin&Rossignac98,Pajarola99].

Finally, while we have presented a lower bound of 2.24 V + O(1) bits for encoding a quad mesh, we have not shown that 2.24 V is the greatest possible lower bound. Future work may lead to a greatest lower bound on the number of possible quad meshes and to more efficient worst-case encodings.

## ACKNOWLEDGMENTS

Thanks to Dietrich Nowottny for providing several finite-element quadrilateral meshes for our tests, and to Greg Turk for providing the original version of the bunny model and the knot and torus models. Thanks to Renato Pajarola for his comments on enumerating quadrilateral meshes, and to Jack Snoeyink for pointing out that CL combinations are possible at valence-two vertices. This work was supported by NSF Grant 9721358.